\documentclass[pre,floatfix,preprint]{revtex4}
\usepackage{epsfig}
\usepackage{amsmath}
\usepackage{amssymb}

\begin{document}
\title{Statistics of defect motion in spatiotemporal chaos in
inclined layer convection}
\author{Karen E. Daniels}
\email{ked@phy.duke.edu}
\altaffiliation[Present address: ]{Department of Physics and Center
for Nonlinear and Complex Systems, Duke University, Durham, NC 27708}
\author{Eberhard Bodenschatz}
\email{eb22@cornell.edu}
\homepage{http://milou.msc.cornell.edu}
\affiliation{Laboratory of Atomic and Solid State Physics, 
Cornell University, Ithaca, NY 14853}
\date{\today}

\begin{abstract} 
We report experiments on defect-tracking in the state of undulation
chaos observed in thermal convection of an inclined fluid layer.
We characterize the ensemble of defect trajectories according to their
velocities, relative positions, diffusion, and gain and loss rates.
In particular, the defects exhibit incidents of rapid transverse 
motion which result in power law distributions for a number of
quantitative measures. We examine connections between this behavior 
and L\'evy flights and anomalous diffusion. In addition, we describe
time-reversal and system size invariance for defect creation and
annihilation rates. 
\end{abstract}

\maketitle



\bigskip 

{\bf 
Topological defects within patterns are observed in many systems to
move in a spatiotemporally chaotic fashion. We examine the motion of
such defects within a defect-turbulent 
state observed in thermal convection of an inclined fluid layer. We
characterize the trajectories of the defects both by analogy to fluid
turbulence -- velocity distributions, diffusion, and power spectra --
and by properties dependent upon the topological characteristics of
the defects: pair creation/annihilation and interactions.
}

\bigskip
\hrule
\bigskip

\section{Introduction}

\begin{figure}
\centerline{\epsfig{file=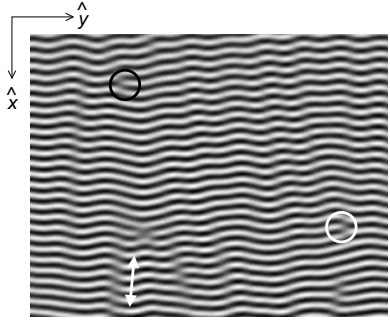, width=2in}}
\caption{Sample Fourier-filtered shadowgraph image of inclined layer
convection at $\epsilon = 0.08$ and $\gamma = 30^\circ$. Black circle
encloses a positive defect; white, a negative. Arrow indicates tearing
region of low-amplitude convection. Uphill direction is at left side
of page. Region shown is the subregion of size $51d \times 63d$ used
for analysis. } 
\label{D_f_pic}
\end{figure}

Nonequilibrium systems with similar symmetries often form patterns
which appear to be universal in spite of having been formed by
different physical mechanisms \cite{Cross:1993:PFE}. In particular, 
reduced descriptions of the patterns often quantify the
similarities in behavior so that understanding of one system can lead 
to insights in multiple systems. A class of spatiotemporally chaotic
states exhibiting defect-mediated turbulence \cite{Coullet:1989:DMT}
has been found in such 
diverse systems as wind-driven sand, electroconvection in liquid
crystals \cite{Rehberg:1989:TWD}, nonlinear optics
\cite{Ramazza:1992:STD}, fluid convection \cite{LaPorta:2000:PMP,
Morris:1993:SDC}, and autocatalytic chemical reactions
\cite{Ouyang:1996:TFS}. In many cases, such systems have been modeled
via the complex Ginzburg-Landau equation \cite{Falcke:1999:SBD,
Haeusser:1997:AMD, Gil:1990:SPD, Granzow:2001:NDU,
Echebarria:2000:DCO}. These various 
defect turbulent patterns are characterized by an underlying striped
state which contains dislocations (point defects) where the stripes
dead-end within the pattern. Locally, the defects distort the
orientation and wavenumber of the stripes and the nucleation,
motion, and annihilation of the defects constitute a
spatiotemporally chaotic system. An example from inclined layer
convection is shown in Fig.~\ref{D_f_pic}.

Previous work on defect turbulence has focused both on snapshots of
such patterns \cite{Gil:1990:SPD, Rehberg:1989:TWD, Ramazza:1992:STD}
and the dynamics and interaction
\cite{Falcke:1999:SBD,Haeusser:1997:AMD,Granzow:2001:NDU}. However,
there are numerous open questions about defect turbulence: 
characterization of the defect motions, interactions between the
defects, and the extent to which the analogy to turbulence is
appropriate. The degree to which such characteristics are similar in
different defect-turbulent systems remains to be explored.

\begin{figure}
\centerline{\epsfig{file=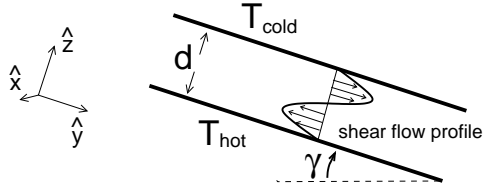,width=2.5in}}
\caption{Schematic of inclined layer convection with associated
coordinate system. $\Delta T \equiv T_\mathrm{hot} - T_\mathrm{cold}$.}
\label{D_f_schematic}
\end{figure}

Investigations of pattern formation in variants of Rayleigh-B\'enard
convection (RBC) have been particularly fruitful
\cite{Bodenschatz:2000:RDR}. The state of undulation chaos (shown in
Fig.~\ref{D_f_pic} and described in \cite{Daniels:2000:PFI,
Daniels:2002:DTI, Daniels:2002:UUC}) observed 
in inclined layer convection (Fig.~\ref{D_f_schematic}) exhibits
defect turbulence and is well suited to 
investigations on the dynamics of defects since 
spatially extended systems and fast time scales are experimentally
accessible. This allows for tracking of point defects through
their creation, motion, and annihilation. In the observed pattern, the
stripes contain  undulations as well as defects; both are
spatiotemporally chaotic (further characterization to be published in
\cite{Daniels:2002:UUC}).  

\begin{figure}
\centerline{\epsfig{file=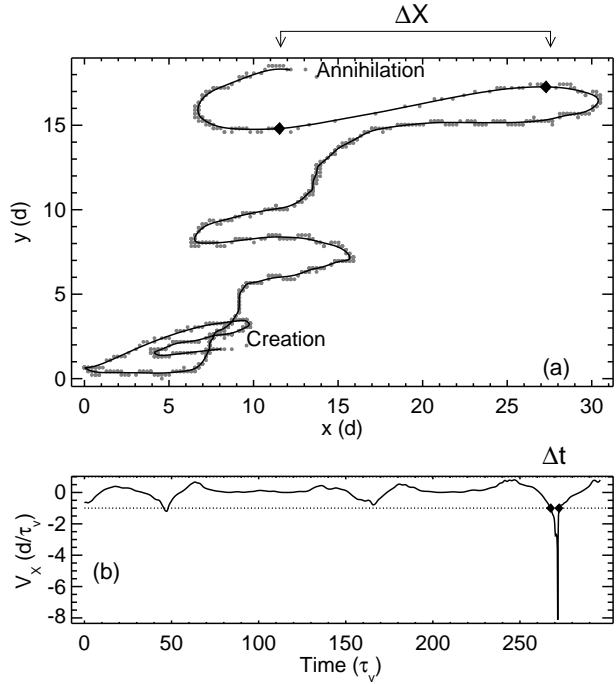, width=3.3in}}
\caption{ (a) Example of a
long trajectory for a defect with positive topological
charge at $\epsilon = 0.08$. Positive $\hat{\bf y}$ indicates a
downslope direction (see Fig.~\protect\ref{D_f_pic}). (b)
Corresponding transverse 
velocity. The symbols $\blacklozenge$ mark the ends of a flight with
displacement $\Delta X$ and duration $\Delta t$. Dotted line is the
trigger velocity for detecting flights.} 
\label{D_f_defecttrack}
\end{figure}

A number of features stand out in the defect trajectories we observe
in undulation chaos. As exemplified in Fig.~\ref{D_f_defecttrack}, the
inherent anisotropy (due to the inclination) is apparent, with the
trajectory meandering 
preferentially in the transverse direction. Occasionally, there is a
rapid burst of transverse 
motion as marked by the black diamonds, corresponding to a tearing of
the pattern
across the rolls. Such behavior appears be related to L\'evy flights
\cite{Klafter:1996:BBM, Shlesinger:1993:SK,
Geisel:1995:LWC} for which distributions of step sizes display power
laws. Furthermore, we are able to study the ensemble of 
trajectories to gain insight into  defect pair creation, interaction, 
and annihilation.

\section{Inclined Layer Convection}

In inclined layer convection (ILC), a thin fluid layer heated from one
side and cooled from the other is tilted by an angle $\gamma$; the
system is anisotropic due to the resulting shear flow (see
Fig.~\ref{D_f_schematic}). The fluid becomes unstable above a critical
temperature difference $\Delta T_c$. At fixed $\gamma$, we describe
how far the system is above the onset of convection via the
non-dimensional driving parameter 
$\epsilon \equiv \frac{\Delta T}{\Delta T_c(\gamma)} - 1$.

At low angles of inclination, buoyancy provides the primary
instability (analogous to RBC) and the
convection rolls are aligned with the shear flow (longitudinal) both
at onset and for the secondary transition to undulation chaos
\cite{Clever:1977:ILC}.  Over a range of intermediate angles 
($15^\circ \lesssim \gamma \lesssim  70^\circ$ for Prandtl number
$\sigma \approx 1$), the longitudinal
rolls become unstable to undulation chaos above $\epsilon \approx
0.01$. It is this 
defect turbulent state which we investigate; some of its properties
have been previously described in other  work
\cite{Daniels:2002:DTI,Daniels:2002:UUC}. Above $\epsilon
\approx 
0.1$, there is a further transition to a state of competing ordered
undulations and undulation chaos. We examine trajectories from both
sides of this transition.

The apparatus used in this experiment is of the type described in de
Bruyn et al.
\cite{deBruyn:1996:ASR}, modified to allow for
inclination. The fluid used was CO$_2$ at a pressure of 
$(56.5 \pm 0.01)$ bar regulated to $\pm 0.005$ bar with a mean
temperature of $(28 \pm 0.05) ^\circ$C regulated to $\pm
0.0003^\circ$C. As determined from a materials properties program 
\cite{deBruyn:1996:ASR}, the Prandtl number was $\sigma = 1.140 \pm
0.001$. A cell of height $d = (388 \pm 2) \mu$m and dimensions $203d
\times 100d$ was used, for which the vertical diffusion time was 
$\tau_v = d^2/\kappa =  (1.532 \pm 0.015)$ sec.  The fluid was weakly
non-Boussinesq conditions: $Q =  0.8$, as described in
\cite{Bodenschatz:2000:RDR} for horizontal fluid layers. All
experiments were performed at a fixed inclination of $\gamma =
30^\circ$, within the regime of buoyancy-instability. Images of the
convection pattern were collected using a digital CCD camera, via the
usual shadowgraph technique \cite{deBruyn:1996:ASR,
Trainoff:2002:POT}. Images were collected at 3 frames per second in
one of two formats.  Six-hour ($ 2 \times 10^4 \tau_v$, 80000 frames)
continuous runs of data were obtained at two values of $\epsilon$:
0.08 (four runs) and $0.17$ (two runs).  For 17 values of $\epsilon$
between 0.04 and 0.22, short runs with 100 images were collected,
separated by at 
least $100 \tau_v$ for statistical independence. At each of these
$\epsilon$, at least $400$ repeats (up to 600 for the lowest values of
$\epsilon$) were performed. Each value of $\epsilon$ was reached by a
quasistatic temperature increase from below. In addition, a run with
quasistatic temperature decreases was performed between $\epsilon =
0.12$ and $\epsilon = 0.06$ to check for  
hysteresis, which was not observed. Only data from the
central homogeneous region of dimension $51d \times 63d$ was utilized
during the analysis unless noted otherwise; see
\cite{Daniels:2002:UUC} for details on the choice of this
region. Size-dependent effects of this region are discussed here.   

\section{Defect Trajectories} 

Here we consider only the defects themselves --- an ensemble of
moving, interacting, charged ``particles'' --- and
not the underlying 
pattern. By reducing the data to a set of defect
trajectories of known charge, birth, and death we can observe how
these particles move individually and as an ensemble. Single defects
may enter or leave through the edges of the system, or be
created/annihilated in pairwise events between defects of opposite
topological charge \cite{Daniels:2002:DTI}. During their lifetimes, 
the defects have positions and velocities which may be related both to
the underlying pattern and to the presence of other defects. We
examine properties averaged over these effects, and also isolate some 
effects due to the latter.

The topological defects are located at points where a roll pair
ends within the pattern (see Fig.~\ref{D_f_pic}). At these points,
there is a phase 
discontinuity and the topological charge is determined
from the phase jump along a contour around the defect: 
\begin{equation}
\oint \vec{\nabla} \phi \cdot d \vec{s} = n 2 \pi,
\label{e_defint}
\end{equation}
The phase field $\phi(x,y)$ is determined from Fourier-demodulated
images for which the complex field $ \psi(x,y) = |A(x,y)| ~ e^{i
\phi(x,y)}$ has been reconstructed. We detect the locations of defects
by finding points where $\Re(\psi)= 0$ and $\Im(\psi) = 0$ and a
phase discontinuity occurs. (Details of this technique are
described in \cite{Daniels:2002:UUC}.) In undulation chaos, we observe  
only defects with $n = \pm 1$; examples of each topological charge
are marked by circles in Fig.~\ref{D_f_pic}.

After the defects were detected in each image, the defect locations
were connected to form trajectories. This was done by matching each
defect already on a trajectory to its closest same-signed
neighbor in the subsequent frame. If the reverse process also agreed
on the match, then the subsequent defect was added to the list of
defects located on that trajectory. Similarly, the ends of positive and
negative defect trajectories were matched up to obtain the locations
of all pair 
creations and annihilations. When the analysis was completed, broken
trajectories (those missing a creation or annihilation) were
rejoined.

Since the defect positions were determined only to the nearest pixel,
the data was later smoothed to interpolate the coordinates along the
trajectory. Each $x_i \equiv x(t_i)$ and $y_i \equiv y(t_i)$ along a
trajectory was replaced with the weighted average of its neighbors,
using the  Gaussian weighting function
\begin{eqnarray}
\nonumber
&w_{ij}  \equiv
\frac{1}{\sigma_x \sigma_y \sigma_t (2 \pi)^{3/2}} \,
\exp[{ - \frac{(x_i-x_j)^2}{2 \sigma^2_x} }] \\ 
&\cdot \exp[{ - \frac{(y_i-y_j)^2}{2 \sigma^2_y} }] \, \cdot \, 
\exp[{ - \frac{(t_i-t_j)^2}{2 \sigma^2_t} }]
\label{D_e_weight}
\end{eqnarray}
This resulted in a smoothed trajectory with coordinates 
$\tilde{x}_i = \sum w_{ij} x_j$
and $\tilde{y}_i = \sum w_{ij} y_j$ (tildes will be dropped
hereafter for convenience). The appropriate widths  
$\sigma$ in $w_{ij}$ were determined by examining a range of
parameters and finding convergence for $\sigma_x = \sigma_y = 2 \, d$
and  $\sigma_t = 2 \, \tau_v$. The advantage of this weighting method
is that it automatically adjusts the fit length so that along
faster-moving segments of the trajectory we average fewer data points
than along slower-moving segments.

From the trajectories $x(t)$ and $y(t)$, we applied a local weighted
linear fit (again using $w_{ij}$) at each point along the trajectory
to obtain the corresponding velocities $v_x(t)$
and $v_y(t)$.
We also determined $v \equiv \sqrt{v_x^2 + v_y^2}$ and 
$\theta \equiv \arctan{\frac{v_x}{v_y}}$, corresponding to a
downslope ($\hat{\bf y}$) direction at $\theta = 0^\circ$. A raw
trajectory is shown in Fig.~\ref{D_f_defecttrack}, along with the smoothed
trajectory and the transverse velocity component $v_x(t)$. 

Defect turbulence in inclined layer convection is
anisotropic, with $v_x$ corresponding to glide 
motion across the rolls and $v_y$ to climb motion along the
rolls. These are fundamentally different behaviors: the former adjusts
the orientation of the rolls, and the later the wavenumber. 
As expected, the decomposed motions are found to be poorly correlated
with each other, with a linear correlation coefficient $R^2 = 0.16$
for positive defect trajectories and $R^2 = 0.12$ for
negative. Therefore, 
we separate all defect positions and motions into their two components 
and examine each direction independently. 

\begin{figure}
\centerline{\epsfig{file=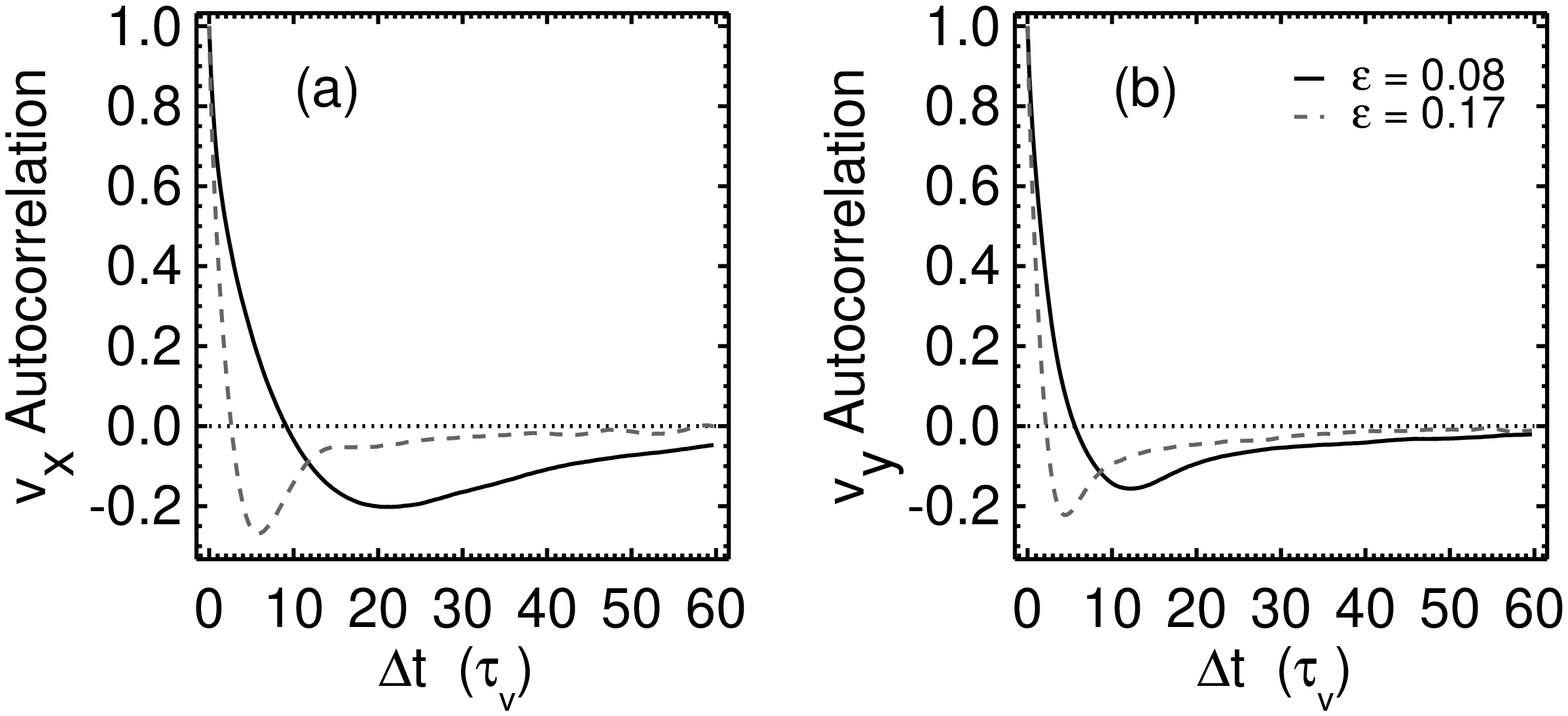, width=3.3in}}
\caption{Velocity autocorrelation in (a) transverse and (b) 
longitudinal directions at $\epsilon = 0.08$ and $\epsilon =
0.17$. Data is shown for positive defects; results for negative defect
trajectories were similar.} 
\label{D_f_velautocorr}
\end{figure}

Defect velocities are observed to be correlated over time scales of
less than $\approx 10 \tau_v$, as determined from temporal
autocorrelation functions of 
$v_x(t)$ and $v_y(t)$. The autocorrelations are plotted in 
Fig.~\ref{D_f_velautocorr},
averaged over the ensemble of trajectories. The defect motions
exhibit short-range order, with exponential decay during
the time before the zero-crossing.

\section{Defect Motion}

We characterize the meanderings of the defect via its central
moments, represented by integer powers of the deviations from its mean
value. For trajectories $x_i(t)$ with mean $\overline{x_i(t)}$, the 
$n^{\mathrm {th}}$ moment is given by the quantity
\begin{equation}
\mu_n(t) \equiv \langle ( x_i(t) - \overline{x_i} ) ^n \rangle
\label{e_moments}
\end{equation}
where $\langle \cdot \rangle$ represents an ensemble average, and a
similar calculation can be made for the $y(t)$. For
normal diffusion or random walks, $\mu_n \propto t^{n/2}$, where
$\mu_2 = 2 D t$ provides a value $D$ for the diffusion 
constant. 

Investigations in a broad variety of situations
\cite[see, for example, ][and references therein]{Bouchaud:1990:ADD,
Geisel:1995:LWC, Klafter:1996:BBM, Shlesinger:1993:SK} have found
anomalous diffusion instead, where $\mu_2 = 2 D t^\alpha$. For $\alpha  
< 1$ a system is said to be subdiffusive, $\alpha > 1$ is 
superdiffusive, and  $\alpha=2$ is ballistic transport. 
Anomalous diffusion is frequently associated with L\'evy
walks/flights, in which the PDFs of the lengths and durations of the
flights are power laws and have infinite moments. Since 
the central limit theorem no longer holds in such cases, the
probability of long jumps (waits) will enhance (retard) the diffusive
behavior. Similarly, the power spectrum $S(\omega)$ of the associated
velocities will exhibit power law behavior as well. 
Some recent examples of such behavior in fluid dynamics include
tokamak density fluctuations \cite{Zaslavsky:2000:LSB} and
particle motion in two-dimensional rotating 
fluid flows \cite{Solomon:1993:OAD}. Defect random walks have   
also been considered for models of intracellular Ca$^{2+}$ dynamics
\cite{Falcke:1999:SBD}, in which subdiffusion was observed. 

\begin{figure}
\centerline{\epsfig{file=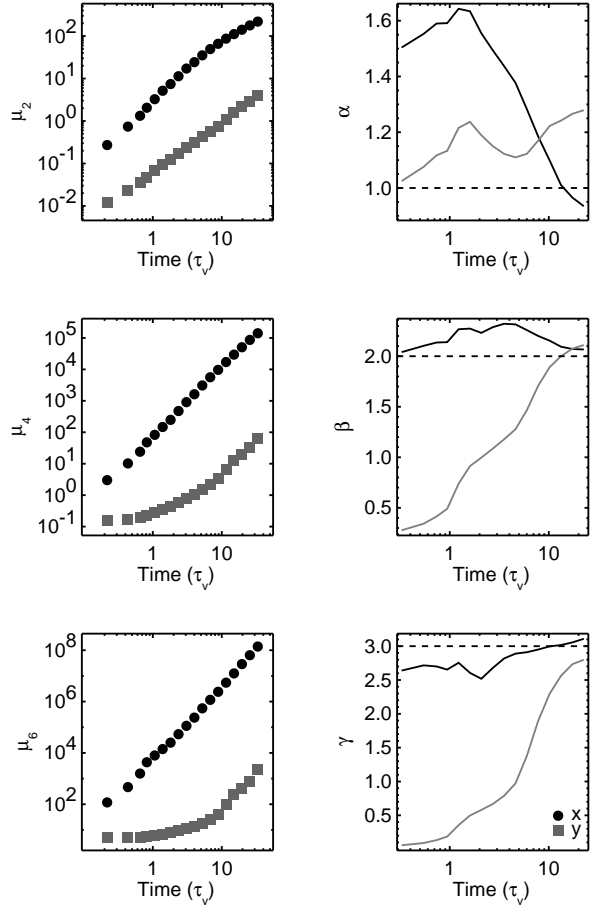, width=3.3in}}
\caption{Displacement moments (Eqn.~\ref{e_moments}) and local fits to
the associated exponents in $\hat{\bf x}$ (black) and $\hat{\bf 
y}$ (gray) directions at $\epsilon = 0.08$. Dashed lines are values
for normal diffusion or random walks, $\mu_n \propto t^{n/2}$. Data is
shown for positive defects; results for negative defects were
similar. }  
\label{D_f_diff8}
\end{figure}

\begin{figure}
\centerline{\epsfig{file=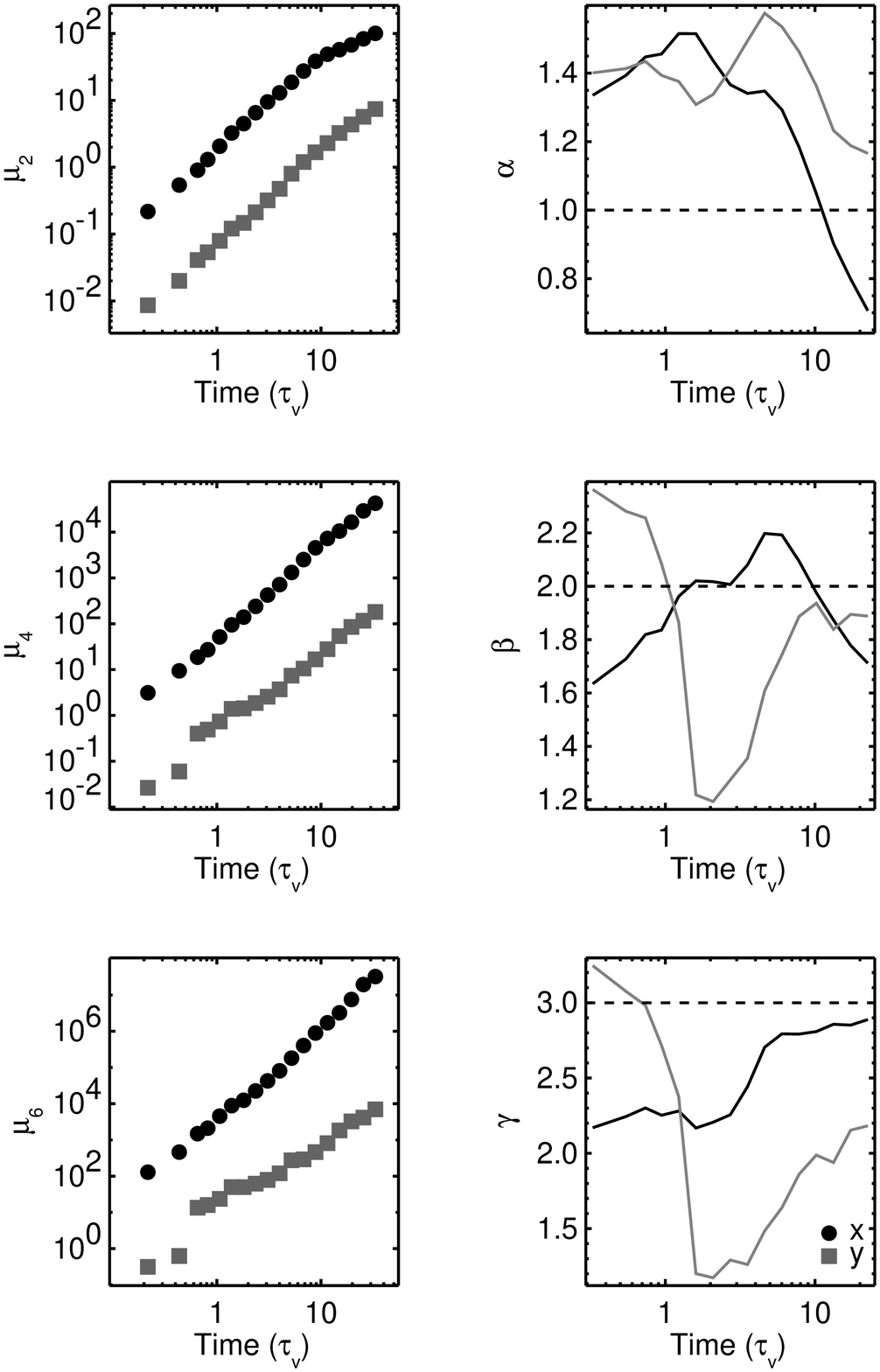, width=3.3in}}
\caption{Displacement moments (Eqn.~\ref{e_moments}) and local fits to
the  associated exponents in $\hat{\bf x}$ (black) and $\hat{\bf 
y}$ (gray) directions at $\epsilon = 0.17$. Dashed lines are values
for normal diffusion or random walks, $\mu_n \propto t^{n/2}$. Data is
shown for positive defects; results for negative defects were
similar. } 
\label{D_f_diff17}
\end{figure}

In Fig.~\ref{D_f_diff8} and \ref{D_f_diff17} we calculate moments for
those values of $t$ at 
which 100 or more trajectories have at least that duration. We look for 
behavior of the type $\mu_2 \propto t^\alpha$, $\mu_4 \propto t^\beta$,
and $\mu_6 \propto t^\gamma$ in both the $\hat{\bf x}$ and $\hat{\bf
y}$ directions. In the variance ($\mu_2$), superdiffusive
behavior is evident at both  $\epsilon$, particularly in  $\hat{\bf 
x}$ where $\alpha \approx 1.5$. The exponents for $\hat{\bf y}$ are 
lower but also indicate superdiffusion.  The exponents of $\mu_4$ and
$\mu_6$ do not suggest 
superdiffusion in $\hat{\bf x}$, and the $\hat{\bf y}$ behavior does
not follow a power law.

\begin{figure}
\centerline{\epsfig{file=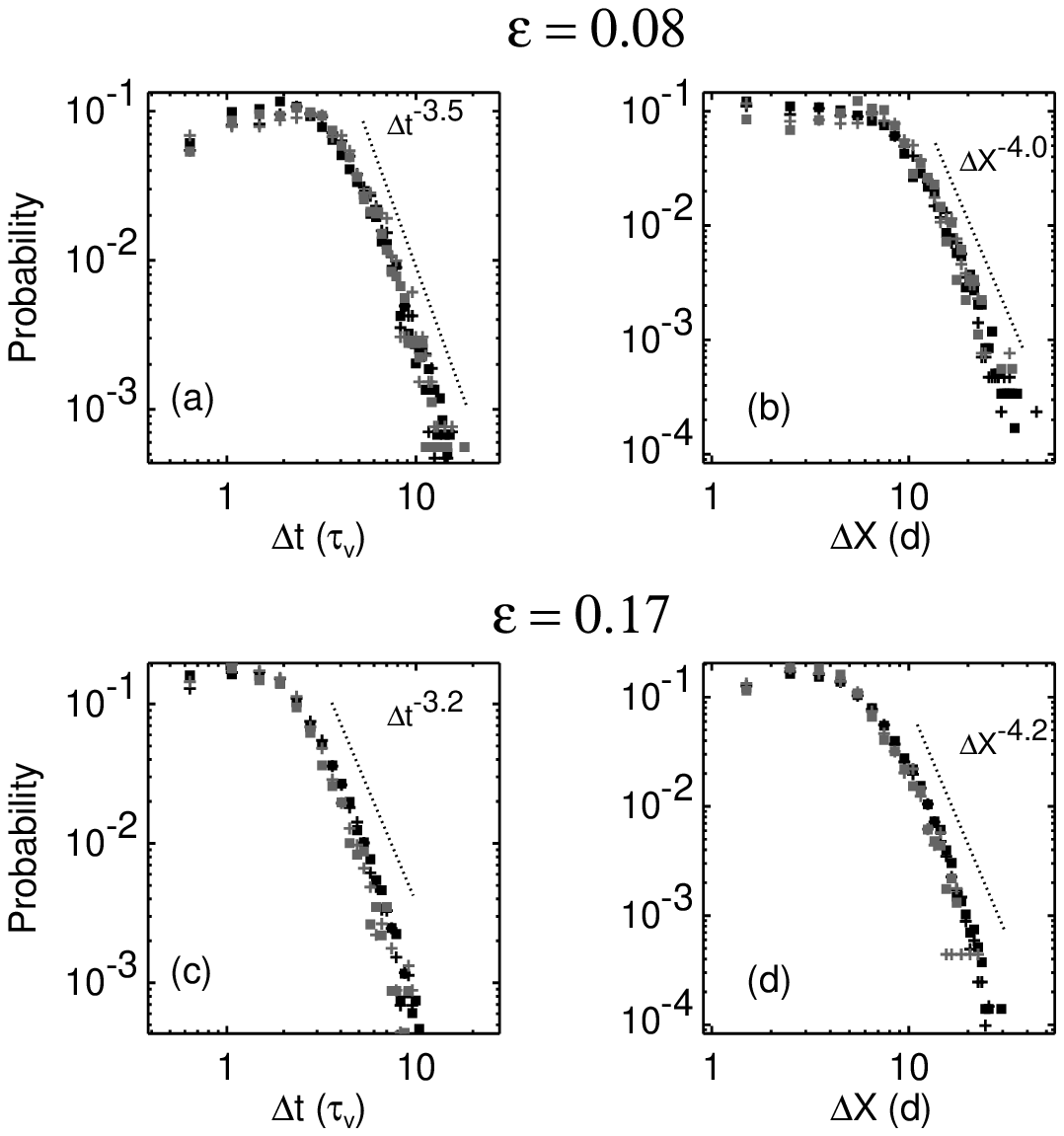, width=3.3in}}
\caption{Distributions of flight (a,c) durations $\Delta t$ and (b,d)
displacements $\Delta X$ at $\epsilon = 0.08$ and $\epsilon =
0.17$. Positive defects are $+$, negative 
defects are squares, black points are from entire trajectories, and   
gray from segments within the homogeneous subregion. Dotted lines
represent fit region for power laws.}
\label{D_f_flights}
\end{figure}

We locate defects flights as the portions of the trajectories which
occur while $|v_x| > 1 \,  d/\tau_v$. An example flight is shown in
Fig.~\ref{D_f_defecttrack}. 
For each of these flights, we determine the duration $\Delta t$ and
displacement $\Delta X$. Distributions of these values are shown in
Fig.~\ref{D_f_flights}. Due to the constraints of a finite system size,
the maximum observable $\Delta X$ is limited.  Nonetheless, the
distribution of flight times $\Delta t$ is observed to have behavior
consistent with a power law tail of exponent approximately $-3$. For
$\Delta X$, behavior suggestive of a power law with exponent
approximately $-4$ is observed, indicating that there is not a
characteristic velocity associated with the flights.  

Physically, these flight events correspond to a tearing of the rolls
when the pattern shifts along a line in the $\hat{\bf x}$ 
direction, so that each roll broken in the process rejoins with the
one next to it.  Such a tearing line is visible next to the white
arrow in  Fig.~\ref{D_f_pic}, where there is weaker convection along
the slopes of the undulations. Since each roll must move by only $2d$
to reconnect with the next roll, defects can be transported
rapidly from one end of the tear line to the other.

\begin{figure}
\centerline{$\epsilon = 0.08$}
\centerline{\epsfig{file=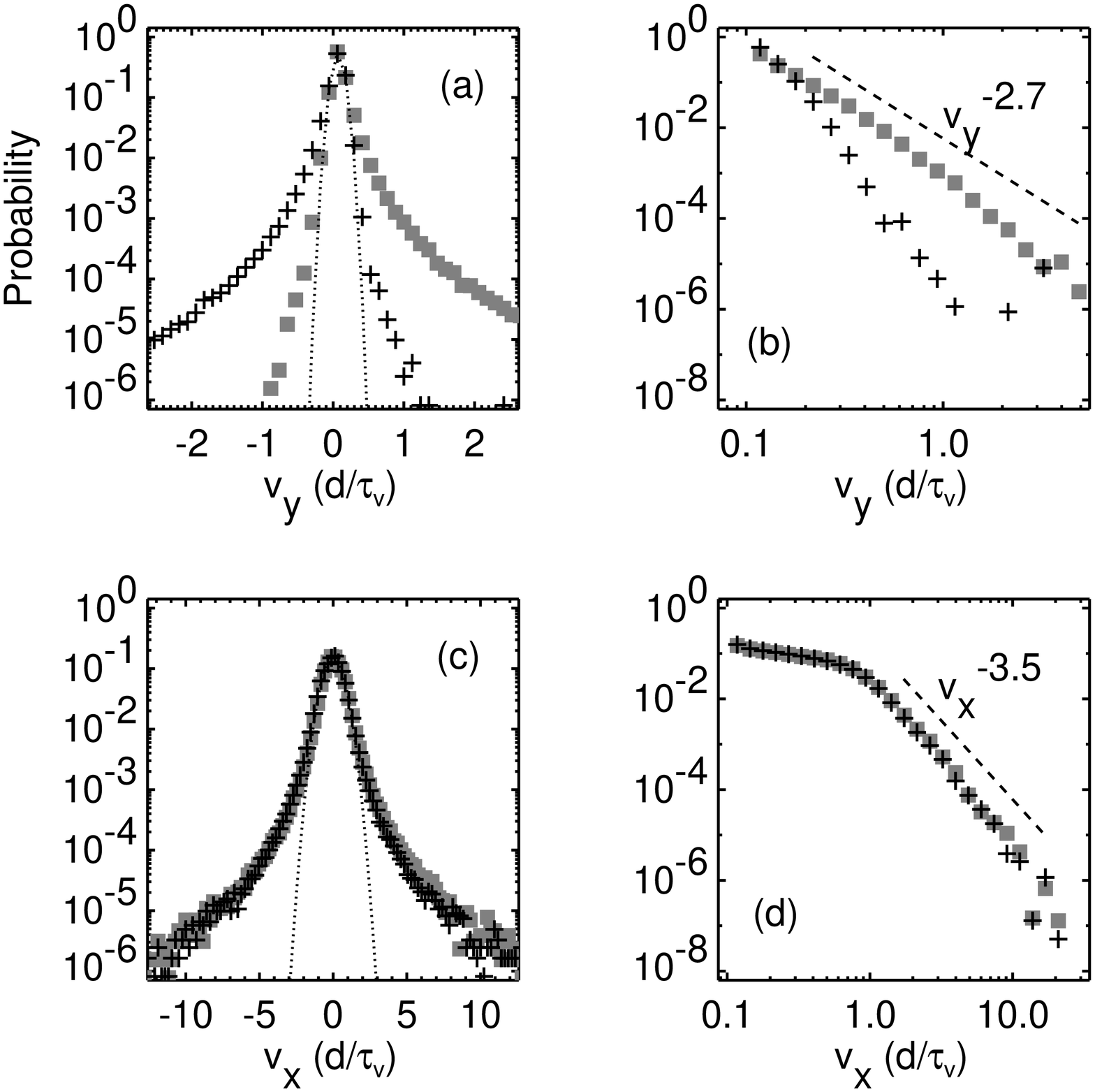, width=3.3in}}
\caption{(a,b) Longitudinal and (c,d) transverse velocity 
PDFs at $\epsilon = 0.08$ on log-log and semi-log (for $v>0$)
axes. Positive defects are shown with black $+$ and negative defects 
with gray squares. Approximately $10^6$ defect velocity measurements
were used to determine the probabilities in each graph. Dotted lines
are Gaussian fits to whole data, with standard deviations $\sigma_y =
0.08$ and $\sigma_x = 0.59$. Dashed lines represent fit region for
power laws. } 
\label{D_f_veldist8}
\end{figure}

\begin{figure}
\centerline{$\epsilon = 0.17$}
\centerline{\epsfig{file=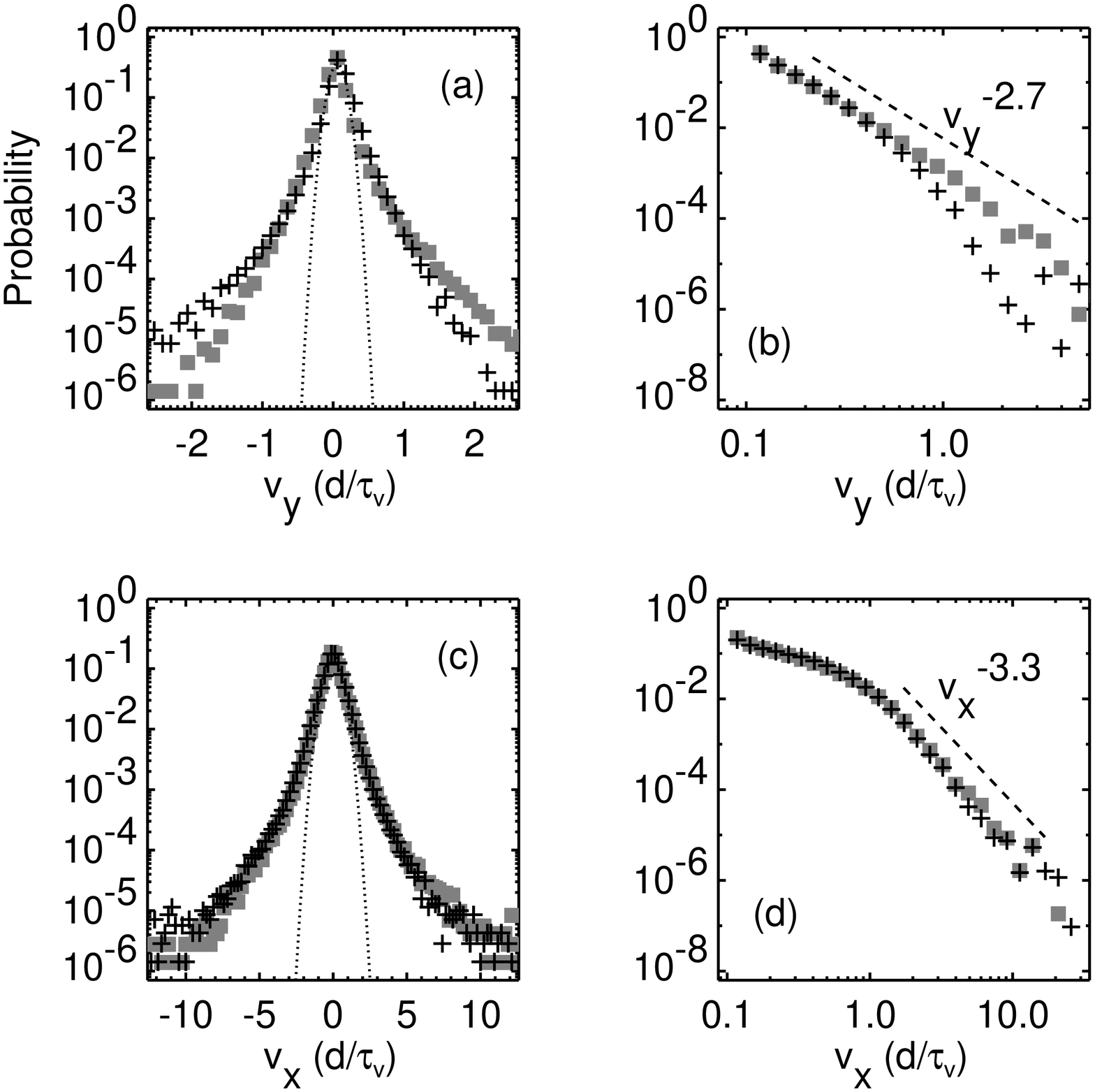, width=3.3in}}
\caption{(a,b) Longitudinal and (c,d) transverse velocity 
PDFs at $\epsilon = 0.17$ on log-log and semi-log (for $v>0$)
axes. Positive defects are shown with black $+$ and negative defects 
with gray squares. Approximately $10^6$ defect velocity measurements
were used to determine the probabilities in each graph. Dotted lines
are Gaussian fits to whole data, with standard deviations $\sigma_y =
0.10$ and $\sigma_x = 0.51$. Dashed lines represent fit region for
power laws. } 
\label{D_f_veldist17}
\end{figure}

\begin{figure}
\centerline{\epsfig{file=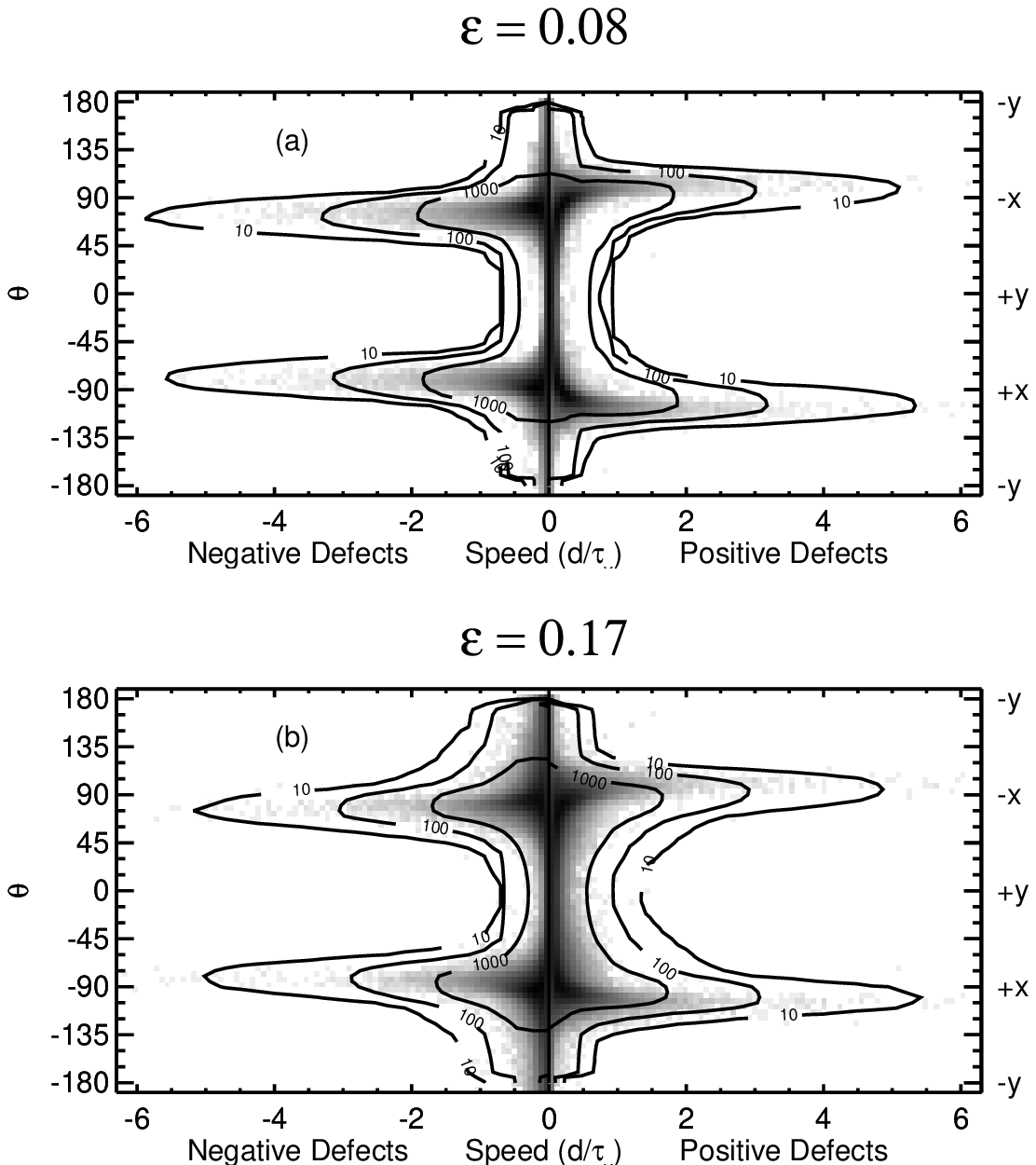, width=3.3in}}
\caption{Histogram for ($v$, $\theta$) pairs. Negative defects plotted 
with negative speed and positive defects plotted with positive
speed. Grayscale is shown on a logarithmic scale. (a) $\epsilon =
0.08$ (b) $\epsilon = 0.17$.}  
\label{D_f_velanglecorr}
\end{figure}

By analogy to fluid turbulence, we plot probability distribution
functions (PDFs) and power spectra of the defect
velocities. Figs.~\ref{D_f_veldist8} and \ref{D_f_veldist17} 
show the results for distributions in $\hat{\bf x}$ and $\hat{\bf y}$.
The PDFs of $v_x$ are independent of topological charge and show steep
power law tails with an exponent of approximately $-3.5$. This is 
consistent with the results obtained for defect flights displacements
and durations.
As can be seen in  Figs.~\ref{D_f_veldist8} and \ref{D_f_veldist17},
the peak of the distributions have a Gaussian shape, indicative of
smoothing due to noise. Each distribution is also of a shape described
by Tsallis statistics of the type described in \cite{Beck:2001:DFN}.
The PDFs for $v_y$ show lower characteristic
velocities and (particularly at $\epsilon=0.08$) dependence on the
topological charge. 
Fig.~\ref{D_f_velanglecorr} shows a complementary picture in
$(v, \theta)$ space via a two dimensional histogram. The $\hat{\bf x}$
flights are now visible as the ridges at 
approximately $\theta = \pm90^\circ$. This transverse direction not only
comprises most of the defect motion, but is also the direction along
which motion is the fastest.

A second effect visible in Figs.~\ref{D_f_veldist8},
\ref{D_f_veldist17}, and \ref{D_f_velanglecorr} is an asymmetry in the
$\pm \hat{\bf y}$ behavior which shifts the $(v,\theta)$ peaks in the
upslope or downslope direction, away from $\pm 90^\circ$. 
Although the $\epsilon=0.17$ defects are more isotropic in  
their motion than the $\epsilon=0.08$ defects, in both cases the
negative defects prefer the downslope direction over the upslope, and
the positive defects prefer the upslope. In fact,
Figs.~\ref{D_f_veldist8} and \ref{D_f_veldist17} show that this
asymmetry is the dominant behavior at lower $\epsilon$: it is
rare for defects to travel other than in the preferred direction.
Asymmetries in defect behavior with respect to topological charge
have been observed in other anisotropic systems such as Langmuir
circulations \cite{Bhaskaran:2002:ELL}, although for quantities such as the
surface convergence rather than the defect velocity. In this case, the
asymmetry may be related to the breaking of the $\hat{\bf z}$ symmetry
by non-Boussinesq effects.

\begin{figure}
\centerline{\epsfig{file=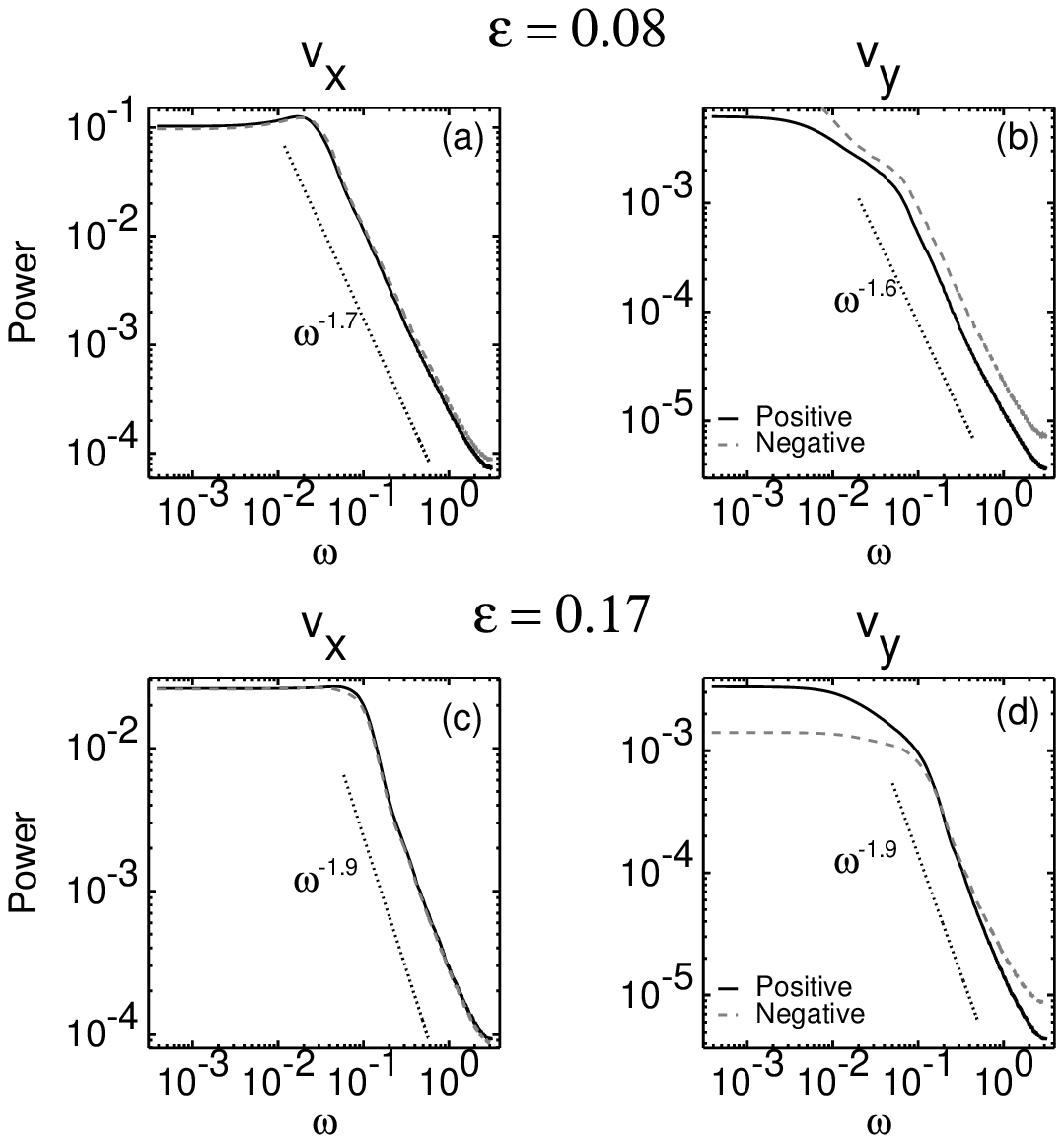, width=3.3in}}
\caption{Power spectra of (a, c) $v_x$ 
and (b, d) $v_y$ at $\epsilon = 0.08$ and 0.17 for positive (black solid)
and negative (gray dashed) defects. Dotted lines show power law fit
region for positive defects.}    
\label{D_f_spectrum}
\end{figure}

In a system consisting of L\'evy flight behavior, the power spectra
$S(\omega)$ of  $v(t)$ is expected to show power law behavior for $\pi
\gg \omega \gg \omega^* = 1/\Delta t^*$, where $\Delta t^*$ is a short
time associated with the low-$t$ limit of power law behavior of
the PDF of $\Delta t$ for flights \cite{Geisel:1995:LWC}. Such a
cutoff is relatively large ($\approx 5 \tau_v$) in the defect flight
durations shown in Fig.~\ref{D_f_flights}, limiting the range over
which we would expect to see power laws in $S(\omega)$ for defect
trajectories. Fig.~\ref{D_f_spectrum}
suggests random-walk 
behavior (constant, uniform spectrum) for low $\omega$, crossing over
to a flight-related power law at high $\omega$.  The exponent in all
cases ($\epsilon$, $\hat{\bf x}$, $\hat{\bf y}$, and topological
charges) is approximately $-1.8$.

\clearpage

\section{Defect Interactions}

\begin{figure}
\centerline{\epsfig{file=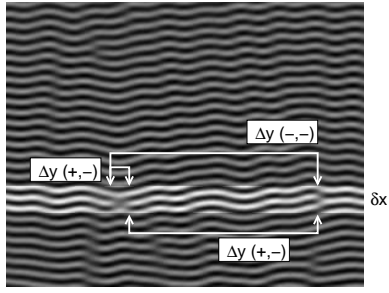, width=2in}}
\caption{Schematic diagram of determination of defect pair separations
in $\hat{\bf y}$, with associated $\Delta y$ measurements.}  
\label{D_f_strippic}
\end{figure}

The defects carry topological charge, and the resulting phase fields
allow them to interact with each other. By examining the relative
position of nearby defects, we can gain insight into their
interactions. We examine strips of width $\delta y =2d$ (or $\delta x$)
and calculate the distances $\Delta x$ (or $\Delta y$)
separating all pairs of defects within the strip 
(see Fig.~\ref{D_f_strippic} for a schematic). 

\begin{figure}
\centerline{\epsfig{file=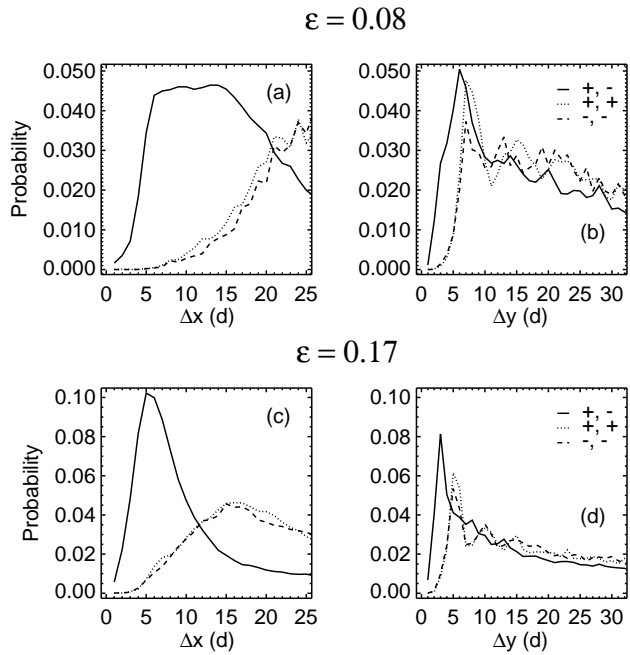, width=3.3in}}
\caption{PDF of defect pair separations ($\Delta x$ and $\Delta y$)
for strips of width $\delta y = \delta x = 2d$ 
as shown in Fig.~\protect\ref{D_f_strippic}. (a,c)
Transverse separations and (b,d) longitudinal separations. Solid lines
are $(+,-)$ pairs, dotted are $(+, +)$, and dashed are
$(-,-)$. Approximately $10^5$ elements used for each curve.}  
\label{D_f_stripsep}
\end{figure}

Fig.~\ref{D_f_stripsep} plots PDFs of this data for all
$(+,-)$, $(+,+)$, and $(-,-)$ pairs of defects: ${\cal P}(\Delta x \,
| \, \delta y < 2d)$ and ${\cal P}(\Delta y \, | \, \delta x <
2d)$. The data is 
limited by the finite size of the region under consideration, and is
only plotted over half the dimensions of the region.
For both $\epsilon$, it is rare for oppositely-charged defects to be
located within $3d$ of each other. This is associated with the
annihilation of such pairs which pass too close in either direction. 

Defect pairs of $(+,+)$ or $(-,-)$ charge behave in similar fashions
to each other, demonstrated by the closely overlapping dashed and
dotted lines of
Fig.~\ref{D_f_stripsep}.  There is a repulsive effect in the
$\hat{\bf x}$ direction, with $\Delta x$ increasing away from 0. The
turnover in the $\epsilon=0.17$ PDF may be due to finite size
considerations.  In the $\hat{\bf y}$ direction, there
is again an excluded region of around $5d$, an effect which is strong 
enough that secondary and tertiary peaks at $10d$ and $15d$ are
visible as well for $\epsilon=0.17$.

\section{Creation and Annihilation Events}

\begin{figure}
\centerline{\epsfig{file=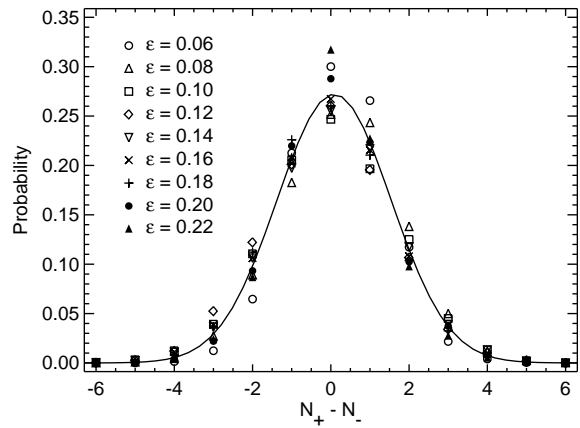, width=3in}}
\caption{PDF of net charge $(N_+ -- N_-)$ found in subregion for nine
values of $\epsilon$. Solid line is 
a Gaussian fit.}
\label{D_f_netcharge}
\end{figure}

A general feature of defect turbulent systems is the production and
loss of defects through two mechanisms: pair creations/annihilations
and flux through the boundaries. A mean-field approximation has been
postulated by Gil et al. \cite{Gil:1990:SPD} and experimentally
examined by Falcke et al. \cite{Falcke:1999:SBD} and Daniels and
Bodenschatz \cite{Daniels:2002:DTI} in which these
rates depend on the number of defects in the system. 
\begin{eqnarray}
C(N) & = & C_0 \\
E(N) & = & E_0 \\
L(N) & = & L_0 N \\
A(N) & = & A_0  N^2
\end{eqnarray}
where $C(N)$ is the probability of a pair creation event happening per 
unit time, $A(N)$ is pair annihilation, $L(N)$ is a single defect
leaving, and $E(N)$ is a single defect entering. $N$ is taken to be
either the 
number of positive defects $N_+$ or the number of negative defects
$N_-$, quantities which are approximately equal on
average. Fig.~\ref{D_f_netcharge} shows PDFs of the net charge $(N_+ -
N_-)$ for the observed images, with a mean of zero (topologically
neutral), independent of $\epsilon$. The constant creation and
entering rates can be understood as being generated by random events,
independent of the number of defects already in the system. The
annihilation rate scales as the number of positive defects times the
number of negative defects. The leaving rate
is proportional to the number of defects present in the system.

Based on this approach, we earlier derived and tested a universal 
distribution for $N$ which agrees with the experimental findings
\cite{Daniels:2002:DTI}. We find, as well,
that the subscripted coefficients depend on $\epsilon$, system size,
and other physical parameters in systematic ways.

\begin{figure}
\centerline{\epsfig{file=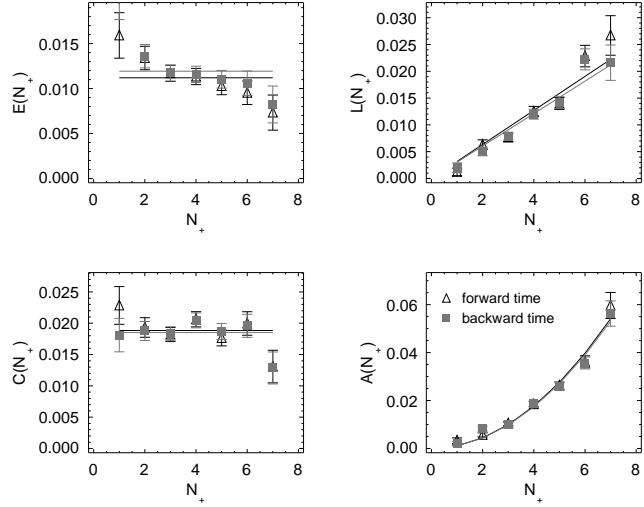, width=3.3in}}
\caption{Comparison of gain/loss coefficients for forward-time
(triangles) and backward-time (squares) trajectories at $\epsilon =
0.07$. Lines are fits to the expected $N$-dependence as described in text.}
\label{D_f_reverse}
\end{figure}

The stationary distribution for $N$ was found using a recursive
relation
\begin{equation}
\text{loss}(N) \, {\cal P}(N) = \text{gain} (N-1) \, {\cal P}(N-1)
\end{equation}
to describe the probabilities at adjacent
$N$ \cite{Daniels:2002:DTI}.  In fact, the
more stringent condition of detailed balance holds as well. By
time-reversing all trajectories, creations become annihilations and
entering defects leave the subregion. Analysis of the reversed
trajectories for $\epsilon = 0.07$ is shown in Fig.~\ref{D_f_reverse}
in comparison with the original data and found to exhibit the same
behavior. 

Such defect gain and loss rates properly scale with the size of
the region under consideration.
We define a set of size-independent coefficients $c_0$,
$a_0$, $e_0$, and $l_0$ to quantify this behavior. Both creation and
entering are random events, which have some rate per unit area and
length, respectively. Thus, $c_0 \equiv \frac{C(N)}{S}$ and $e_0
\equiv \frac{E(N)}{P}$  
where $S$ and $P$ are the surface area and perimeter of the
subregion, respectively. With primed and
unprimed variables representing different subregion sizes,
$c_0 \equiv \frac{C_0 }{S} = \frac{C^\prime_0}{S^\prime}$ and  
$e_0 \equiv \frac{E_0}{P} = \frac{E^\prime_0}{P^\prime}$. For two 
subregions with the same density of defects $n \equiv N/S$, the
annihilations per unit area (at constant $n$) is also constant. For 
$A(N) = A_0 N^2 = A_0 n^2 S^2 = A^\prime_0 n^2 S^{\prime2}$ 
and constant density of defects $n$, 
$a_0 \equiv \frac{A(N)}{S} \propto A_0 S$. 
Analogously, $l_0 \equiv L_0 P$.

\begin{figure}
\centerline{\epsfig{file=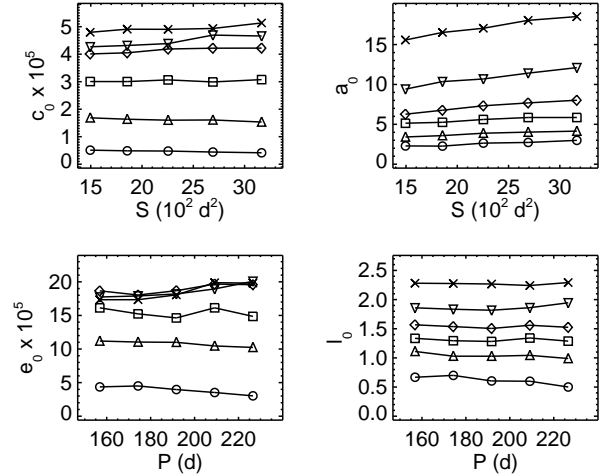, width=3.3in}}
\caption{Size-independent gain/loss coefficients {\it vs.} size of
test region. $\epsilon$ from bottom to top are: 0.06 
($\bigcirc$), 0.09 ($\bigtriangleup$), 0.12 ($\Box$), 0.15 ($\Diamond$), 
0.18 ($\bigtriangledown$), 0.22 ($\times$). Data is shown
for positive defects; results for negative defects were
similar. } 
\label{D_f_coeff_boxsize}
\end{figure}

Fig.~\ref{D_f_coeff_boxsize} compares these rescaled 
coefficients for various sizes of test regions. Because of the
$\hat{\bf x}$-$\hat{\bf y}$ anisotropy, regions with the same aspect
ratio as the homogeneous subregion were used.  All four coefficients
show constant rescaled rates for a given $\epsilon$. There is,
however,  a slight trend in the annihilation rates:
larger boxes allow more defects to find each other
and annihilate, particularly at higher $\epsilon$. This may also
relate to the larger test regions approaching the less homogeneous
sidewall regions (see \cite{Daniels:2002:UUC}).

\begin{figure}
\centerline{\epsfig{file=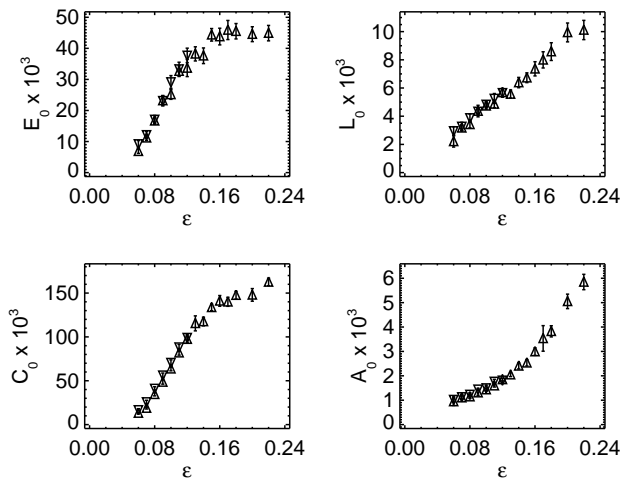, width=3.3in}}
\caption{Rate coefficients as a function of $\epsilon$ for homogeneous
subregion. $\bigtriangleup$ represent data taken under quasistatic 
temperature increase; $\bigtriangledown$ represent data taken under
quasistatic temperature decrease. }
\label{D_f_coeffs}
\end{figure}

Fig.~\ref{D_f_coeffs} shows the coefficients $C_0$, $E_0$, $L_0$, and
$A_0$ as a function of $\epsilon$. The error bars
were determined via a bootstrap  method: by
re-sampling the data with replacement we obtained a distribution of
values for each coefficient from which to estimate the error. At
present we have no 
explanation for the particular shapes of these graphs, which appear to
be linear to first order. Both $C_0(\epsilon)$ and $E_0(\epsilon)$
appear to intercept the $\epsilon$-axis close to the $\epsilon_c
\approx 0.02$ onset of 
undulation chaos \cite{Daniels:2002:DTI}. For $A_0(\epsilon)$ and
$L_0(\epsilon)$, 
both lines extrapolate to the origin, the expected behavior for
$N=0$. The data at $\epsilon = 0.04$ and 0.05 have been
disregarded due to finite-size effects which are significant at low
values of $\epsilon$ since the undulations are a long-wavelength
($k=0$) instability. Defects are increasingly rare for $\epsilon
\rightarrow \epsilon_c$ as the wavelength of the undulations becomes 
on the order of the size of the convection cell.

\section{Conclusion}

The defect trajectories observed in inclined layer convection display
many intriguing behaviors. The occasional rapid motion of defects
across many convection rolls can be associated with power laws in
various quantitative measures of their motility: diffusion, velocity
PDFs, flight size PDFs, and velocity power spectra. While the system
is strongly anisotropic and its defect motions are dominated by
transverse motion, PDFs of the defect separations reveal significant
correlations in the longitudinal direction as well. Finally, the
trajectories allow us to probe the gain and loss of defects in which
much of this interesting behavior can be averaged over to reveal
general results. 

We have focused primarily on characterizing the observed flight
behavior. If these rapid motions are, in fact, related to L\'evy
flights then the exponents found in each of the power laws should be
related to each other as has been presented in the literature for
various random walk formalisms \cite{Geisel:1995:LWC,
Shlesinger:1993:SK, Weeks:1998:ADR}. However, comparisons to
existing theory are difficult due to both the strongly anisotropic
nature and the finite size of this system. Further investigation into
this phenomenon, possibly in other systems, will undoubtedly be
fruitful. 

The trajectories analyzed here were obtained for $\epsilon = 0.08$
and 0.17, of which the former is in a state of undulation chaos and the
later intermittently exhibit ordered undulations. The dynamics of this
transition is discussed in \cite{Daniels:2002:UUC}, and the defect 
trajectories should 
also be examined in light of the existence of the two states. The
statistics presented here will presumably vary according to whether
the system is in an ordered 
or disordered state. Furthermore, the motion of the defects during the
transition may shed light on the nature of the transition. 

Other questions remain regarding the relation between the defect
motion and the underlying undulation pattern. For instance, the
relationship between the motion of the defects and the stability of
the local wavenumber. A related issue is that defects have been
observed to ``bounce'' off regions of ordered undulations, sharply
reversing direction. 

Finally, the relationship between the relative position and relative 
velocity of defects has not yet been investigated. Further work in
this area would provide information about the attraction and repulsion
of defects, particularly near creation and annihilation events.

We are grateful to G. M. Zaslavksy, T. Geisel, J. P. Sethna, C. L. 
Henley, C. Beck, and N. Mordant for useful discussions and the NSF for 
support under DMR-0072077.

\end{document}